\documentclass[10pt, twoside]{article}

\usepackage[tbtags]{amsmath}
\usepackage{accents}
\usepackage{rotating}
\usepackage{amsfonts}
\usepackage{amssymb}
\usepackage{latexsym}

\DeclareMathSymbol{\vecarrow}{\mathord}{letters}{"7E}

\newlength{\lvech}
\newlength{\lvecw}

\newcommand{\starMC}{*_{\scriptscriptstyle MC}}
\newcommand{\starM}{*_{\scriptscriptstyle M}}
\newcommand{\starC}{*_{\scriptscriptstyle C}}

\newcommand{\fett}[1]{\mbox{\boldmath$#1$}} %selbsterklärend
\newcommand{\fetts}[1]{\mbox{\scriptsize\boldmath$#1$}} %selbsterklärend
\newcommand{\bsigma}{\fett{\sigma}} % bold sigma
\newcommand{\bssigma}{\fetts{\sigma}} % bold sigma
\newcommand{\vsigma}{\vec\bsigma}
\newcommand{\vssigma}{\vec\bssigma}
\newcommand{\beq}{\begin{equation}}
\newcommand{\eeq}{\end{equation}}

\newcommand{\iu}{\mathrm{i}}

\newcommand{\bra}[1]{\left<{#1}\right|}
\newcommand{\ket}[1]{\left|{#1}\right>}
\newcommand{\braket}[2]{\left<{#1}\;\vrule\;{#2}\right>}
\newcommand{\braketop}[3]{\left<{#1}\;\vrule\;{#2}\;\vrule\;{#3}\right>}

\begin{document}
\makeatletter
\title{Spin Description in the Star Product and the Path Integral Formalism}
\author{
S. Odendahl\footnote{sven.odendahl@uni-dortmund.de} \, and \,
P. Henselder\footnote{peter.henselder@uni-dortmund.de} \\
Fachbereich Physik, Universit\"at Dortmund\\
44221 Dortmund}

\maketitle

\begin{abstract}
The spin can be described in the star product formalism by extending
the bosonic Moyal product in the fermionic sector.  The fermionic
star product is then the Clifford product of geometric algebra and
it is possible to formulate the fermionic star product formalism in analogy to
the bosonic star product formalism. For the fermionic star product
description of spin, one can then establish the relation to other
approaches that describe spin with fermionic variables, i.e. the
operator formalism and the path integral formalism. It is shown that
the fermionic star product formalism and the fermionic path integral
formalism are related in analogy to their bosonic counterparts.
\end{abstract}

%%%%%%%%%%%%%%%%%%%%%%%%%%%%%%%%%%%%%%%%%%%%%%%%%%%%%%%%%%%%%%%%%%%%%%%%%%%%%%%%%%%%%%%%%%%%%%%%%%%%%%%%%%%%%%%%%%%%%%%%%%%%%%%%%%%%%%%%%%%%%%%%%
\section{Introduction}
%%%%%%%%%%%%%%%%%%%%%%%%%%%%%%%%%%%%%%%%%%%%%%%%%%%%%%%%%%%%%%%%%%%%%%%%%%%%%%%%%%%%%%%%%%%%%%%%%%%%%%%%%%%%%%%%%%%%%%%%%%%%%%%%%%%%%%%%%%%%%%%%%
\qquad The star product formalism as it was established  by Bayen et
al. in \cite{Bayen} gives an alternative description of quantum mechanics
on the phase space \cite{Zachos}. One might then wonder if it is
also possible to describe spin in the star product formalism.
Berezin and Marinov showed in \cite{Berezin} how the spin can be
described in the context of pseudoclassical mechanics. So it appears
natural to apply the program of deformation quantization to
pseudoclassical mechanics in order to obtain a description of spin
with a fermionic star product \cite{Deform3}. With the fermionic
star product it is then possible to find in analogy to the bosonic
star product formalism spin Wigner functions as eigenfunctions of a
fermionic star eigenvalue equation and a spin star exponential that
describes the time development.

A fermionic star product already appeared in \cite{Bayen}, whereas a
systematic account to a fermionic star product formalism was given in
\cite{Mexiko}. With the fermionic star product the Grassmann algebra
is deformed into a Clifford algebra, so that such a Clifford star product
can be interpreted as the geometric product of geometric algebra in a
superanalytic formulation \cite{Deform5}. Geometric algebra goes
back to early ideas of Hamilton, Grassmann, and
Clifford and was developed into a full formalism by Hestenes in
\cite{Hestenes1} and \cite{Hestenes3}. From then on it was applied
to a wide range of physical problems and in particular it was used
for the description of spin (for a thorough discussion of geometric
algebra see \cite{Doran1}). The superanalytic formulation of
geometric algebra and its relation to pseudoclassical mechanics was
established in \cite{Doran2} so that - in conjunction with the
fermionic Clifford star product - a natural geometric interpretation
of the fermionic star product formalism is obtained. Furthermore, it
now appears natural to combine the fermionic and the bosonic star
product formalism (see for example \cite{Fradkin}). 
This leads then to a noncommutative version of geometric algebra,
where the noncommutativity leads to the natural appearance of a 
spin term \cite{Deform5} and for geometric algebra on the phase 
space to a natural appearance of supersymmetric quantum mechanics
\cite{SusyGA}.

In the next section we will first give a short overview of the formal
structures in the bosonic star product formalism so that one can
compare this to the fermionic case. In the third section we will
then give a short introduction to geometric algebra in its
superanalytic formulation with a Clifford star product. Thereafter,
we show how the spin arises naturally from noncommutative
geometric algebra and how it can also be formulated in two isomorphic
Clifford algebras. Furthermore, we establish the connection of the
spin description by spin projectors or spin Wigner functions to the
spinor description in \cite{Doran1}.  In section six and seven we
then describe how the spin can be formulated with a fermionic
operator and a fermionic path integral formalism. For both approaches
the relation to the fermionic star product formalism is established.
In particular we show that the relation of the fermionic path integral
and the fermionic star product formalism is analogous to the
bosonic one that was found in \cite{Cohen,Sharan}.

%%%%%%%%%%%%%%%%%%%%%%%%%%%%%%%%%%%%%%%%%%%%%%%%%%%%%%%%%%%%%%%%%%%%%%%%%%%%%%%%%%%%%%%%%%%%%%%%%%%%%%%%%%%%%%%%%%%%%%%%%%%%%%%%%%%%%%%%%%%%%%%%%
\section{The Bosonic Star Product Formalism}
\setcounter{equation}{0}\label{BFosc}
%%%%%%%%%%%%%%%%%%%%%%%%%%%%%%%%%%%%%%%%%%%%%%%%%%%%%%%%%%%%%%%%%%%%%%%%%%%%%%%%%%%%%%%%%%%%%%%%%%%%%%%%%%%%%%%%%%%%%%%%%%%%%%%%%%%%%%%%%%%%%%%%%
\qquad On the $2d$-dimensional flat phase space with coordinates
$(q^i,p^i)$ the Moyal product of two phase space functions $f(q,p)$ and
$g(q,p)$ is defined as
\begin{equation}
f\starM g = f\exp\left[\frac{\iu\hbar}{2}\sum_{i=1}^d
\left(\frac{\overleftarrow{\partial}}{\partial q^i}\frac{\overrightarrow{\partial}}{\partial p^i}-
\frac{\overleftarrow{\partial}}{\partial p^i}\frac{\overrightarrow{\partial}}{\partial q^i}\right)\right]g.
\label{starMDef}
\end{equation}
It can also be written in integral form, which in a two-dimensional
phase space reads \cite{Baker}
\begin{equation}
f\starM g=\frac{1}{\hbar^2\pi^2}\int dq'dq''dp'dp'' f(q',p')g(q'',p'') \exp\left(\frac{2}{\iu\hbar}
\big(p(q'-q'')+p'(q''-q)+p''(q-q')\big)\right).
\label{starMInt}
\end{equation}
The star product replaces the conventional product between
functions on the phase space and it is so constructed that the
star anticommutator corresponds to the Poisson bracket:
\begin{equation}
\lim_{\hbar\rightarrow 0}\frac{1}{\iu\hbar}
\left[f,g\right]_{\starM}=\lim_{\hbar\rightarrow 0}
\frac{1}{\iu\hbar}\left(f\starM g-g\starM
f\right) =\{f,g\}_{PB}.
\end{equation}
This relation is the principle of correspondence. The states of a
system with Hamilton function $H(q,p)$ are described by Wigner
functions $\pi_n^{(M)}(q,p)$. The Wigner functions and the
corresponding energy levels $E_n$ can be calculated with the help of
the star exponential $\mathrm{Exp}_M(Ht)(q,p)$, which is defined as
\begin{equation}
\mathrm{Exp}_M(Ht)=e_{\starM}^{-\frac{\iu t}{\hbar}H}=
\sum_{n=0}^{\infty}\frac{1}{n!}\left(\frac{-\iu t}{\hbar} \right)^n
H^{n\starM}= \sum_{n=0}^{\infty}\pi_n^{(M)}e^{-\iu E_nt/\hbar},
\label{ExpDef}
\end{equation}
where $H^{n\starM}=H\starM\cdots\starM H$ is the $n$-fold star
product of $H$. The star exponential fulfills the analogue of the
time dependent Schr\"{o}dinger equation
\begin{equation}
\iu\hbar\frac{d}{dt}\mathrm{Exp}_M(Ht)=H\starM\mathrm{Exp}_M(Ht)
\label{ExpSchroe}
\end{equation}
and describes the time development of a phase space function
$f(q,p)$ \cite{Zachos}:
\begin{equation}
f(t)=\overline{\mathrm{Exp}_M(Ht)}\starM f\starM\mathrm{Exp}_M(Ht).
\end{equation}
The connection of the star exponential and the path integral was
established in \cite{Cohen,Sharan}. It was shown that the path
integral is the Fourier transform of the star exponential, i.e.
\begin{equation}
\int \frac{DqDp}{2\pi\hbar}\, \exp\left[\frac{\iu}{\hbar}\int_{t_i}^{t_f}(p\dot{q}-H)dt\right]
=\int\frac{dp}{2\pi\hbar}\exp\left[\frac{\iu}{\hbar}p(q_f-q_i)\right]
\mathrm{Exp}_M(Ht)\left((q_f+q_i)/2,p\right).
\end{equation}
Putting now (\ref{ExpDef}) into (\ref{ExpSchroe}) gives the
$\starM$-eigenvalue equation
\begin{equation}
H\starM\pi_n^{(M)}= E_n\pi_n^{(M)}\label{Schroe}
\end{equation}
and for $t=0$ these equations lead to the spectral decomposition
of the Hamilton function:
\begin{equation}
H=\sum_{n=0}^{\infty}E_n\pi_n^{(M)}.
\end{equation}
Substituting this expression in (\ref{Schroe}) gives
\begin{equation}
\pi_n^{(M)}\starM\pi_m^{(M)}=\delta_{mn}\pi_n^{(M)},
\end{equation}
which together with the completeness relation $\sum_{n=0}^{\infty}
\pi_n^{(M)}=1$, that follows from (\ref{ExpDef}) for $t=0$, means
that the Wigner functions are projectors.

%%%%%%%%%%%%%%%%%%%%%%%%%%%%%%%%%%%%%%%%%%%%%%%%%%%%%%%%%%%%%%%%%%%%%%%%%%%%%%%%%%%%%%%%%%%%%%%%%%%%%
\section{Geometric Algebra and the Clifford Star Product}
\setcounter{equation}{0}
%%%%%%%%%%%%%%%%%%%%%%%%%%%%%%%%%%%%%%%%%%%%%%%%%%%%%%%%%%%%%%%%%%%%%%%%%%%%%%%%%%%%%%%%%%%%%%%%%%%%%
In the supersymmetric formulation of geometric algebra, the
basisvectors of a $d$-dimensional vector space are Grassmann
variables $\fett{\sigma}_i$, $i=1,\ldots,d$. A general multivector
is then a Grassmann number
\begin{equation}
A=A^0+A^i\fett{\sigma}_i+A^{i_1i_2}\fett{\sigma}_{i_1}\fett{\sigma}_{i_2}
+\cdots +A^{i_1\cdots i_d}\fett{\sigma}_{i_1}\cdots\fett{\sigma}_{i_d}.
\end{equation}
On this vector space one has the fermionic Clifford star product,
which can be derived from the pseudo-classic formalism for fermions
in various ways as was shown in \cite{Berezin}, \cite{Casalbuoni},
or \cite{Deform3}:
\begin{equation}
A\starC B=A\,\exp\left[\eta_{ij}\frac{\overleftarrow{\partial}}{\partial
\fett{\sigma}_i} \frac{\overrightarrow{\partial}}{\partial\fett{\sigma}_j}
\right]\,B, \label{starCDef2}
\end{equation}
where $\eta_{ij}$ is the metric. In three euclidian dimensions for
example, the general multivector has a scalar, a vector, a bivector
and a pseudoscalar part, i.e.\ $A=\langle A\rangle_0 +\langle
A\rangle_1+\langle A\rangle_2+\langle A\rangle_3$, where
$\langle\;\rangle_n$ projects on the term of Grassmann grade $n$. In
the euclidian case the Clifford star product can just as the Moyal
product be written in integral form. One has
\begin{equation}
A\starC B=\int d^3\bsigma'd^3\bsigma''A(\vsigma')B(\vsigma'')
\exp\left[\sum_{i=1}^3\left(\bsigma_i\bsigma'_i+\bsigma'_i\bsigma''_i+\bsigma''_i\bsigma_i\right)\right],
\label{starCInt}
\end{equation}
where the integral $\int d^3\bsigma=\int d\bsigma_3\int
d\bsigma_2\int d\bsigma_1$ is the Berezin integral \cite{Berezin}
and $\vsigma$ is the tuple $(\bsigma_1,\bsigma_2,\bsigma_3)$ so that
$A(\vsigma')$ and $B(\vsigma'')$ are the multivectors $A$ and $B$
with $\bsigma_i$ substituted by ${\bsigma'_i}$ and ${\bsigma''_i}$
respectively. All these sets of Grassmann variables are mutually
anticommuting. The Clifford star product of two basis vectors is
given by $\bsigma_i\starC\bsigma_j =\bsigma_i\bsigma_j+\delta_{ij}$,
so that for two vectors $\fett{a}=a^i\bsigma_i$ and
$\fett{b}=b^i\bsigma_i$ the scalar and the exterior product can be
defined as the symmetric and the antisymmetric part of the Clifford
star product:
\begin{equation}
\frac{1}{2}\{\fett{a},\fett{b}\}_{\starC}=\fett{a}\cdot\fett{b}
\quad\mathrm{and}\quad
\frac{1}{2}\left[\fett{a},\fett{b}\right]_{\starC}=\fett{a}\fett{b}
\equiv\fett{a}\wedge\fett{b},
\end{equation}
with the star anticommutator $\{A,B\}_{\starC}=A\starC B+B\starC A$
and the star commutator $\left[A,B\right]_{\starC}=A\starC B-B\starC A$.
The scalar and the exterior product can be generalized to homogenous
multivectors $A_{(m)}$ and $B_{(n)}$ of Grassmann grade $m$ and $n$
as $A_{(m)}\cdot B_{(n)}=\langle A_{(m)}\starC B_{(n)}\rangle_{|m-n|}$ and
$A_{(m)}B_{(n)}=\langle A_{(m)}\starC B_{(n)}\rangle_{m+n}$.

In three dimensions there are furthermore three basis bivectors
\begin{equation}
\mathtt{B}_i=I_{(3)}\starC \fett{\sigma}_i,
\end{equation}
where $I_{(3)}=\fett{\sigma}_1\fett{\sigma}_2\fett{\sigma}_3$ is the
unit pseudoscalar in three dimensions. For the basis bivectors one has
\begin{equation}
\frac{1}{2}\{\mathtt{B}_i,\mathtt{B}_j\}_{\starC}=-\delta_{ij}
\quad\mathrm{and}\quad
\frac{1}{2}\left[\mathtt{B}_i,\mathtt{B}_j\right]_{\starC}=-\varepsilon_{ijk}
\mathtt{B}_k.
\end{equation}
And in the basis $\mathtt{i}=\mathtt{B}_1$, $\mathtt{j}=-\mathtt{B}_2$
and $\mathtt{k}=\mathtt{B}_3$ the bivectors fulfill
\begin{equation}
\mathtt{i}^{2\starC}=\mathtt{j}^{2\starC}=\mathtt{k}^{2\starC}
=\mathtt{i}\starC\mathtt{j}\starC\mathtt{k}=-1,
\end{equation}
so that the even multivector $Q=Q^0+Q^1\mathtt{i}+Q^2\mathtt{j}
+Q^3\mathtt{k}$ is a quaternion.

The bivectors generate the rotations \cite{Doran3}, i.e.\ a rotation of the vector
$\fett{x}$ around the axis that is given by the unit vector $\fett{n}$ is
generated by the bivector $\mathtt{B}=I_{(3)}\starC\fett{n}$ and the
rotated vector $\fett{x}'$ is given by
\begin{equation}
\fett{x}'=R\starC \fett{x}\starC \overline{R},
\end{equation}
where $R$ is the rotor, i.e.\ the fermionic star exponential
\begin{equation}
R=e_{\starC}^{\mathtt{B}\varphi/2},
\end{equation}
and the involution reverses the order of the Grassmann variables, so
that $\overline{\mathtt{B}}=-\mathtt{B}$ and, in addition, the
scalar coefficients are complex conjugated. A multivector is said to
be real if $\overline{A}=A$.

While a vector transforms with a left and a right action of the rotor, a
spinor $\psi$ transforms with a left action, i.e.
\begin{equation}
\psi'=R\starC\psi.
\end{equation}
In three dimensions such a spinor can be represented as an even multivector
\cite{Francis}
\begin{equation}
\psi=\psi^0+\psi^i\mathtt{B}_i=(\psi^0+\psi^3\mathtt{B}_3)
+(\psi^2+\psi^1\mathtt{B}_1)\starC\mathtt{B}_2,\label{psihatdef}
\end{equation}
which has the tuple analogon
\begin{equation}
\hat{\psi}=\left(
\begin{array}{c}
  \psi^0+\iu\psi^3 \\
  -\psi^2+\iu\psi^1 \\
\end{array}
\right).
\end{equation}
Specifically for the up and down spinors one has $\psi_+=1$ and
$\psi_-=\mathtt{B}_2$. With this choice for the spinor the
eigenvalue equation reads
\begin{equation}
\lambda\psi=\bsigma_i\starC\psi\starC\bsigma_3.
\end{equation}
Furthermore $\psi$ as an even multivector can be
interpreted as a not-normalized rotor \cite{Doran1}.

%%%%%%%%%%%%%%%%%%%%%%%%%%%%%%%%%%%%%%%%%%%%%%%%%%%%%%%%%%%%%%%%%%%%%%%%%%%%%%%%%%%%%%%%%%%%%%%%%%%%%
\section{The fermionic star product formalism}
\setcounter{equation}{0}
%%%%%%%%%%%%%%%%%%%%%%%%%%%%%%%%%%%%%%%%%%%%%%%%%%%%%%%%%%%%%%%%%%%%%%%%%%%%%%%%%%%%%%%%%%%%%%%%%%%%%
It is now natural to combine the fermionic Clifford product and the
bosonic Moyal product to obtain a noncommutative version of
geometric algebra. If the coefficients of two vectors do not commute,
the square of a vector is in general not a scalar, but one obtains
in addition to the scalar a bivector term. For the Hamiltonian of a
particle in a homogenous magnetic field with vector potential
\begin{equation}
\fett{A}=\frac{1}{2}\left(B_2q_3-B_3q_2\right)\bsigma_1+
\frac{1}{2}\left(B_3q_1-B_1q_3\right)\bsigma_2+
\frac{1}{2}\left(B_1q_2-B_2q_1\right)\bsigma_3
\end{equation}
so that
\begin{equation}
\fett{\nabla}\fett{A}=\left(\frac{\partial}{\partial
q_1}\bsigma_1+\frac{\partial}{\partial
q_2}\bsigma_2+\frac{\partial}{\partial
q_3}\bsigma_3\right)\fett{A}=\sum_{i=1}^3B_i\mathtt{B}_i,
\end{equation}
one gets
\begin{eqnarray}
H&=&\frac{1}{2m}\Big[\left(p_1+eA_1\right)\fett{\sigma}_1+
\left(p_2+eA_2\right)\fett{\sigma}_2+
\left(p_3+eA_3\right)\fett{\sigma}_3\Big]^{2\starMC}\label{Hqua1}\\
&=&\frac{1}{2m}\sum_{i=1}^3\left(p_i+eA_i\right)^{2\starM}+
\sum_{i=1}^3\epsilon_{ikl}\frac{\hbar\omega_i}{4\iu}\bsigma_k\bsigma_l\\
&=&H_0+\mathtt{H}_S\label{Hqua2}
\end{eqnarray}
with $\omega_i=\frac{eB_i}{m}$. The first term $H_0$ describes the
Landau problem, which can be solved in the star product formalism as
described in \cite{Landau,Cliffordization}. The additional bivector
valued term $\mathtt{H}_S$ describes the spin and one can then in
analogy to the Moyal product formalism obtain star exponentials and
Wigner functions for the spin. The $\starC$-eigenvalue equation reads
\begin{equation}
\mathtt{H}_S\starC\pi^{(C)}_\pm=E\pi^{(C)}_\pm.\label{starCeigenvalue}
\end{equation}
With the assumption that the $\starC$-square of the spin term in the
Hamiltonian is a positive real value, i.e.
\begin{equation}
E^2=\mathtt{H}_S \starC \mathtt{H}_S
\in\mathbb{R}_+,\label{HmustbeReal}
\end{equation}
which applies not only to all real bivectors but also to all real
vectors and some multivectors with vector and bivector part, the
$\starC$-eigenvalue equation (\ref{starCeigenvalue}) is solved by
\begin{equation}
\pi^{(C)}_\pm=\frac{1}{2}\left(1\pm \frac{\mathtt{H}_S}{|E|}\right)
\end{equation}
with $\starC$-eigenvalues $E_\pm=\pm|E|$. It is easy to show that these
solutions fulfill the projector properties
$\pi^{(C)}_\pm\starC\pi^{(C)}_\pm=\pi^{(C)}_\pm$,
$\pi^{(C)}_+\starC\pi^{(C)}_-=\pi^{(C)}_-\starC\pi^{(C)}_+=0$, and
$\pi^{(C)}_++\pi^{(C)}_-=1$. As in the bosonic star product
formalism, the star-exponential describes the time development and
is defined similarly:
\begin{equation}
\mathrm{Exp}_C(\mathtt{H}_St)=e_{\starC}^{-\iu\mathtt{H}_St}=\cos(|E|t)-\iu\frac{\mathtt{H}_S}{|E|}\sin(|E|t).
\end{equation}
It can also be expressed by the Fourier-Dirichlet expansion:
\begin{equation}
\mathrm{Exp}_C(\mathtt{H}_St)=\sum_E\pi^{(C)}_Ee^{-\iu Et}.
\end{equation}

For the specific example of a magnetic field in $z$-direction
with vector potential
$\fett{A}=-\frac{B}{2}q_2\bsigma_1+\frac{B}{2}q_1\bsigma_2$ and
Hamiltonian
$\mathtt{H}_{S_z}=\frac{\hbar\omega}{2\iu}\bsigma_1\bsigma_2$ this
results in
\begin{eqnarray}
\pi^{(C)}_\pm&=&\frac{1}{2}(1\mp \iu\bsigma_1\bsigma_2) \\
\mathrm{and}\qquad\mathrm{Exp}_C(\mathtt{H}_{S_z}t)&=&\cos\left(\frac{\hbar\omega}{2}t\right)-\bsigma_1\bsigma_2\sin\left(\frac{\hbar\omega}{2}t\right).
\end{eqnarray}
With the star exponential it is possible to describe the time
development of the generators $\bsigma_i$ as
\begin{subequations}
\begin{eqnarray}
\bsigma_1(t)&=&\overline{\mathrm{Exp}_C(\mathtt{H}_{S_z}t)}\starC\bsigma_1\starC\mathrm{Exp}_C(\mathtt{H}_{S_z}t)=\bsigma_1\cos(\hbar\omega t)-\bsigma_2\sin(\hbar\omega t),\\
\bsigma_2(t)&=&\overline{\mathrm{Exp}_C(\mathtt{H}_{S_z}t)}\starC\bsigma_2\starC\mathrm{Exp}_C(\mathtt{H}_{S_z}t)=\bsigma_1\sin(\hbar\omega t)+\bsigma_2\cos(\hbar\omega t),\\
\bsigma_3(t)&=&\overline{\mathrm{Exp}_C(\mathtt{H}_{S_z}t)}\starC\bsigma_3\starC\mathrm{Exp}_C(\mathtt{H}_{S_z}t)=\bsigma_3.
\end{eqnarray}
\end{subequations}
The lowering and raising operator for this case can be derived
easily with
$\bsigma_1\starC\pi^{(C)}_\pm\starC\bsigma_1=\pi^{(C)}_\mp$. Using
the projector properties of the Wigner function and defining
$\fett{f}=\pi^{(C)}_+\starC\bsigma_1=\frac{1}{2}(\bsigma_1+\iu\bsigma_2)$
and
$\overline{\fett{f}}=\bsigma_1\starC\pi^{(C)}_+=\frac{1}{2}(\bsigma_1-\iu\bsigma_2)$
this results in
\begin{eqnarray}
\overline{\fett{f}}\starC\pi^{(C)}_+\starC\fett{f}&=&\pi^{(C)}_-\\
\mathrm{and} \qquad
\fett{f}\starC\pi^{(C)}_-\starC\overline{\fett{f}}&=&\pi^{(C)}_+.
\end{eqnarray}
Alternatively
$\fett{f}'=\pi^{(C)}_+\starC\bsigma_2=\frac{1}{2}(\bsigma_2-\iu\bsigma_1)$
and
$\overline{\fett{f}}'=\bsigma_2\starC\pi^{(C)}_+=\frac{1}{2}(\bsigma_2+\iu\bsigma_1)$
could be chosen.

Due to the fact that the spin is described by bivector-valued terms,
the ladder operators should also be elements of the even subalgebra
${\cal C}\ell_3^+(\mathbb{C})$. The obvious choice is to multiply
with $\bsigma_3$ as it commutes with the Wigner functions and one
gets
\begin{eqnarray}
\mathtt{f}&=&\bsigma_3\starC\fett{f}=\frac{1}{2}(\bsigma_3\bsigma_1-\iu\bsigma_2\bsigma_3)\label{RLoperator1}\\
\mathrm{and}\qquad
\overline{\mathtt{f}}&=&\overline{\fett{f}}\starC\bsigma_3=-\frac{1}{2}(\bsigma_3\bsigma_1+\iu\bsigma_2\bsigma_3)\label{RLoperator2}
\end{eqnarray}
or, alternatively,
$\mathtt{f}'=-\frac{1}{2}(\bsigma_2\bsigma_3+\iu\bsigma_3\bsigma_1)$
and
$\overline{\mathtt{f}}'=\frac{1}{2}(\bsigma_2\bsigma_3-\iu\bsigma_3\bsigma_1)$
respectively so that
\begin{eqnarray}
\overline{\mathtt{f}}\starC\pi^{(C)}_+\starC\mathtt{f}&=&\pi^{(C)}_-\\
\mathrm{and}\qquad\mathtt{f}\starC\pi^{(C)}_-\starC\overline{\mathtt{f}}&=&\pi^{(C)}_+.
\end{eqnarray}

%%%%%%%%%%%%%%%%%%%%%%%%%%%%%%%%%%%%%%%%%%%%%%%%%%%%%%%%%%%%%%%%%%%%%%%%%%%%%%%%%%%%%%%%%%%%%%%%%%%%%
\section{Isomorphic Clifford algebras}
\setcounter{equation}{0}
%%%%%%%%%%%%%%%%%%%%%%%%%%%%%%%%%%%%%%%%%%%%%%%%%%%%%%%%%%%%%%%%%%%%%%%%%%%%%%%%%%%%%%%%%%%%%%%%%%%%%
There are two isomorphic Clifford algebras to ${\cal
C}\ell_3^+(\mathbb{C})$. The first algebra is the complex algebra
with only two Grassmann variables ${\cal C}\ell_2(\mathbb{C})$. One
gets the isomorphism by substituting $-\iu\bsigma_2\bsigma_3$ with
$\bsigma_1$ and $-\iu\bsigma_3\bsigma_1$ with $\bsigma_2$. Then the
spin can be described with the elements of ${\cal
C}\ell_2(\mathbb{C})$ in almost the same manner. The Hamiltonian,
the Wigner-function, the star exponential, and the ladder operators
for the spin in $z$-direction are still the same whereas the
Hamiltonian for the spin in $x$- and $y$-direction is
$\fett{H}_{S_x}=\frac{\hbar\omega}{2}\bsigma_1$ and
$\fett{H}_{S_y}=\frac{\hbar\omega}{2}\bsigma_2$ respectively so that
there are with regard to the star product three anticommuting
variables $\bsigma_1$, $\bsigma_2$, and $-\iu\bsigma_1\bsigma_2$.
These can also be formulated in terms of the ladder operators
$\fett{f}$ and $\overline{\fett{f}}$ from the previous chapter as
\begin{subequations}
\begin{eqnarray}
\bsigma_1&=&\fett{f}+\overline{\fett{f}}\\
\bsigma_2&=&-\iu\left(\fett{f}-\overline{\fett{f}}\right)\\
-\iu\bsigma_1\bsigma_2&=&2\fett{f}\overline{\fett{f}}=2\left(\fett{f}\starC\overline{\fett{f}}-1\right)
\end{eqnarray}
\end{subequations}
and thus the Wigner functions for the spin in $z$-direction can be written as
\begin{subequations}
\begin{eqnarray}
\pi^{(C)}_+&=&\fett{f}\starC\overline{\fett{f}}\\
\text{and}\qquad\pi^{(C)}_-&=&\overline{\fett{f}}\starC\fett{f}.
\end{eqnarray}
\end{subequations}
Therefore, these ladder operators correspond to the holomorphic
variables in the bosonic case. This description was for example used
in \cite{Mauro}. Since this is a linear transformation in
$\bsigma_1$ and $\bsigma_2$, integral formulations can be translated
by using the substitution rule for Grassmann variables
\cite{Doran1}:
\begin{equation}
\int
d\bsigma_2\bsigma_1=\det\left|\frac{\partial\vsigma'}{\partial\vsigma}\right|\int
d\bsigma'_2\bsigma'_1.
\end{equation}
Additionally, these are just the operators that are used as creation
and annihilation operators in coherent state formalisms (see for example
\cite{Mexiko}).

Up to now the spin was described using complex geometric algebras,
but the spinors as described in \cite{Doran1} are even elements of
the real geometric algebra ${\cal C}\ell_3(\mathbb{R})$. However,
this is the second algebra that is isomorphic to ${\cal
C}\ell_3^+(\mathbb{C})$ by simply substituting the imaginary unit
$\iu$ with the pseudoscalar $I_{(3)}$ (and consequently
$-\iu\bsigma_1\bsigma_2$ with $\bsigma_3$ and so forth). In this
algebra the spin term of the Hamiltonian for a homogenous magnetic
field in $z$-direction is then a vector
$\fett{H}_{S_z}=\frac{\hbar\omega}{2}\bsigma_3$, but as the assumption
for the $\starC$-eigenvalue equation (\ref{HmustbeReal}) still holds, the
solution is
\begin{equation}
\pi^{(C)}_\pm=\frac{1}{2}\left(1\pm\frac{\fett{H}_{S_z}}{|E|}\right)=\frac{1}{2}(1\pm\bsigma_3).\label{RealWigner}
\end{equation}
The definition of the star exponential changes accordingly:
\begin{eqnarray}
\mathrm{Exp}_C(\fett{H}_{S_z}t)&=&e_{\starC}^{-I_{(3)}\starC\fetts{H}_{S_z}t}=\cos(|E|t)-I_{(3)}\starC\frac{\fett{H}_{S_z}}{|E|}\sin(|E|t)\\
&=&\sum_E\pi^{(C)}_E\starC e_{\starC}^{-I_{(3)} Et}.
\end{eqnarray}

Now there is a direct connection between the Wigner function
$\pi^{(C)}$ for an arbitrary spin direction and the spinor $\psi$.
If $\psi$ is normalized, i.e.
\begin{equation}
\overline{\psi}\starC\psi=\psi\starC\overline{\psi}=1,
\end{equation}
the spinor is simply the rotor which rotates the spin-up Wigner
function so that $\pi^{(C)}$ can be defined as
\begin{equation}
\pi^{(C)}=\psi\starC\frac{1}{2}(1+\bsigma_3)\starC\overline{\psi}.\label{ArbitraryWigner}
\end{equation}
With this definition the general properties of the Wigner function
still have to be met. First, the idempotence of the Wigner-function
is shown:
\begin{eqnarray}
\pi^{(C)}\starC\pi^{(C)}&=&\psi\starC\frac{1}{2}(1+\bsigma_3)\starC\overline{\psi}\starC\psi\starC\frac{1}{2}(1+\bsigma_3)\starC\overline{\psi} \\
    &=&\psi\starC\frac{1}{4}(1+\bsigma_3)\starC(1+\bsigma_3)\starC\overline{\psi} \\
    &=&\psi\starC\frac{1}{2}(1+\bsigma_3)\starC\overline{\psi} \\
    &=&\pi^{(C)}.
\end{eqnarray}
To prove the equivalence of the eigenvalue equations in the two
formalisms, it is necessary to verify the implication in both
directions. With simple algebraic transformations the first
direction yields
\begin{eqnarray}
&&\lambda\psi=\bsigma_i\starC\psi\starC\bsigma_3 \\
\Rightarrow &&\lambda\psi\starC\frac{1}{2}(1+\bsigma_3)\starC\overline{\psi}=\bsigma_i\starC\psi\starC\bsigma_3\starC\frac{1}{2}(1+\bsigma_3)\starC\overline{\psi} \\
\Rightarrow &&\lambda\psi\starC\frac{1}{2}(1+\bsigma_3)\starC\overline{\psi}=\bsigma_i\starC\psi\starC\frac{1}{2}(1+\bsigma_3)\starC\overline{\psi} \\
\Rightarrow &&\lambda\pi=\bsigma_i\starC\pi^{(C)}
\end{eqnarray}
and the other direction
\begin{eqnarray}
&&\lambda\pi^{(C)}=\bsigma_i\starC\pi^{(C)} \\
\Rightarrow &&\lambda\psi\starC(1+\bsigma_3)\starC\overline{\psi}=\bsigma_i\starC\psi\starC(1+\bsigma_3)\starC\overline{\psi} \\
\Rightarrow &&\lambda\psi+\lambda\psi\starC\bsigma_3=\bsigma_i\starC\psi+\bsigma_i\starC\psi\starC\bsigma_3 \\
\Rightarrow &&\lambda\psi+\lambda\psi\starC\bsigma_3-\bsigma_i\starC\psi\starC\bsigma_3-\bsigma_i\starC\psi\starC\bsigma_3\starC\bsigma_3=0 \\
\Rightarrow &&(\lambda\psi-\bsigma_i\starC\psi\starC\bsigma_3)\starC(1+\bsigma_3)=0 \\
\Rightarrow &&\lambda\psi=\bsigma_i\starC\psi\starC\bsigma_3.
\end{eqnarray}
Choosing again the spin in $z$-direction and inserting $\psi_\pm$
from (\ref{psihatdef}) into (\ref{ArbitraryWigner}) yields the
Wigner functions from (\ref{RealWigner}) as expected. The raising
and lowering operator for these functions resulting from
(\ref{RLoperator1}) and (\ref{RLoperator2}) using the isomorphism
are
\begin{eqnarray}
f&=&\frac{1}{2}(\bsigma_1+I_{(3)}\starC\bsigma_2)\\
\mathrm{and}\qquad
\overline{f}&=&\frac{1}{2}(\bsigma_1-I_{(3)}\starC\bsigma_2),
\end{eqnarray}
which is the Wigner function (\ref{RealWigner}) multiplied with
$\bsigma_1$ (depending on the direction from which $\bsigma_1$ is
multiplied and whether $\pi_+$ or $\pi_-$ is chosen one gets $f$ or
$\overline{f}$). So these multivectors essentially reflect the Wigner
function at the $\bsigma_1$-axis and project onto either the up- or
down-spin:
\begin{eqnarray}
\overline{f}\starC\pi^{(C)}_+\starC f&=\pi^{(C)}_-\starC\bsigma_1\starC\pi^{(C)}_+\starC\bsigma_1\starC\pi^{(C)}_-=&\pi^{(C)}_-\\
f\starC\pi^{(C)}_-\starC\overline{f}&=\pi^{(C)}_+\starC\bsigma_1\starC\pi^{(C)}_-\starC\bsigma_1\starC\pi^{(C)}_+=&\pi^{(C)}_+.
\end{eqnarray}

%%%%%%%%%%%%%%%%%%%%%%%%%%%%%%%%%%%%%%%%%%%%%%%%%%%%%%%%%%%%%%%%%%%%%%%%%%%%%%%%%%%%%%%%%%%%%%%%%%%%%
\section{Fermionic operator formalism}
\setcounter{equation}{0}
%%%%%%%%%%%%%%%%%%%%%%%%%%%%%%%%%%%%%%%%%%%%%%%%%%%%%%%%%%%%%%%%%%%%%%%%%%%%%%%%%%%%%%%%%%%%%%%%%%%%%
Going back to the complex algebra with three Grassmann variables,
the Clifford star product can also be expressed by an operator
product by substituting $\bsigma_i$ with
$\hat{\bsigma}_i=\left(\bsigma_i+\frac{\vec{\partial}}{\partial\bsigma_i}\right)$
on the left side of the product. This results in the same
Hamiltonian as in \cite{Lopez}
\begin{equation}
\hat{\mathtt{H}}_S=-|E|\iu\left(\bsigma_1+\frac{\vec{\partial}}{\partial\bsigma_1}\right)\left(\bsigma_2+\frac{\vec{\partial}}{\partial\bsigma_2}\right)
\end{equation}
and shows that the fermionic wave function and the Wigner function
are proportional because
$\hat{\mathtt{H}}_S\psi=\mathtt{H}_S\starC\psi$. The normalization
is determined by a newly defined scalar-product
\begin{equation}
\int d^3\bsigma \tilde\psi(\vsigma)\psi(\vsigma)=\int
d\bsigma_3d\bsigma_2d\bsigma_1\tilde\psi(\vsigma)\psi(\vsigma)=1,
\end{equation}
where $\tilde\psi(\vsigma)$ is defined as
\begin{equation}
\tilde\psi(\vsigma)=\hat{I}_{(3)}\overline\psi(\vsigma)=I_{(3)}\starC\overline\psi(\vsigma)
\end{equation}
so that
\begin{equation}
\int d^3\bsigma
\tilde\psi(\vsigma)\psi(\vsigma)={\left(\psi^0\right)}^2+{\left(\psi^1\right)}^2+{\left(\psi^2\right)}^2+{\left(\psi^3\right)}^2
+{\left(\psi^{12}\right)}^2+{\left(\psi^{23}\right)}^2+{\left(\psi^{31}\right)}^2+{\left(\psi^{123}\right)}^2.
\end{equation}
This yields the following fermionic wave function:
\begin{equation}
\psi_\pm(\vsigma)=\sqrt{2}\;\pi^{(C)}_\pm=\frac{1}{\sqrt{2}}(1\mp\iu\bsigma_1\bsigma_2).
\end{equation}
In Dirac notation this can be written as
\begin{subequations}
\begin{eqnarray}
\braket{\vsigma}{\psi}&=&\psi(\vsigma) \\
\text{and}\qquad\braket{\psi}{\vsigma}&=&\tilde\psi(\vsigma)
\end{eqnarray}\label{PsiDirac}
\end{subequations}
so that
\begin{equation}
\braket{\psi'}{\psi}=\int{d^3\bsigma\;\tilde\psi'(\vsigma)\psi(\vsigma)}=\int{d^3\bsigma\;\braket{\psi'}{\vsigma}\braket{\vsigma}{\psi}},
\end{equation}
which means that
\begin{equation}
\int{d^3\bsigma\;\ket{\vsigma}\bra{\vsigma}}=\hat{1}.\label{UnitDirac}
\end{equation}
With the Fourier-transform as defined in \cite{Berezin} but with
slightly reordered factors
\begin{eqnarray}
F(\vsigma')&=&\int d^3\bsigma f(\vsigma)e^{\iu\vssigma\vssigma'}  \\
f(\vsigma)&=&-\iu\int
d^3\bsigma'F(\vsigma')e^{\iu\vssigma'\vssigma},
\end{eqnarray}
where $\vsigma\vsigma'=\sum_{i=1}^3\bsigma_i\bsigma_i'$, and the
analogous definition of the delta-function
\begin{equation}
f(\vsigma')=\int d^3\bsigma
\delta^3(\vsigma'-\vsigma)f(\vsigma),\label{DeltaDef}
\end{equation}
the Fourier-transform of the delta-function is
$e^{\iu\vssigma'\vssigma}$. The delta function itself can be derived
easily:
\begin{eqnarray}
\delta^3(\vsigma'-\vsigma)&=&-\iu\int d^3\bsigma'' e^{\iu\vssigma''(\vssigma-\vssigma')} \\
&=&\int d^3\bsigma'' e^{\vssigma''(\vssigma'-\vssigma)}\label{DeltaInt}\\
&=&(\bsigma_1-\bsigma_1')(\bsigma_2-\bsigma_2')(\bsigma_3-\bsigma_3').
\end{eqnarray}
In addition, it can be noted using (\ref{starCInt}) and
(\ref{DeltaInt}) that within an integral the star product, which is
the same as the operator product, is identical to the conventional
product, i.e.
\begin{eqnarray}
\int d^3\bsigma \hat{A}(\vsigma)B(\vsigma)&=& \int d^3\bsigma
A(\vsigma)\starC B(\vsigma)\\
&=& \int d^3\bsigma
d^3\bsigma'd^3\bsigma''A(\vsigma')B(\vsigma'')e^{(\vssigma\vssigma'+\vssigma'\vssigma''+\vssigma''\vssigma)}\\
&=&\int d^3\bsigma'd^3\bsigma'' d^3\bsigma
e^{\vssigma\left(\vssigma'-\vssigma''\right)}A(\vsigma')B(\vsigma'')
e^{\vssigma'\vssigma''} \\
&=&\int
d^3\bsigma'd^3\bsigma''\delta^3(\vsigma'-\vsigma'')A(\vsigma')B(\vsigma'')
e^{\vssigma'\vssigma''} \\
&=&\int d^3\bsigma'A(\vsigma')B(\vsigma')e^{\vssigma'\vssigma'} \\
&=&\int d^3\bsigma A(\vsigma)B(\vsigma),
\end{eqnarray}
just as in the bosonic case, where $\int dqdp\, f(q,p)\starM g(q,p)
=\int dqdp\, f(q,p)g(q,p)$. With these definitions and results a possible
choice for the bras and kets is
\begin{alignat}{3}
\bra{\psi}=&\int d^3\bsigma'\tilde\psi(\vsigma') & \ket{\psi}=&\psi(\vsigma') \\
\bra{\vsigma}=&\int d^3\bsigma' \delta^3(\vsigma-\vsigma') \qquad &
\ket{\vsigma}=&-\delta^3(\vsigma-\vsigma'),
\end{alignat}
where the bras are integration operators and all but $\ket{\vsigma}$
are even multivectors. This results furthermore in
\begin{eqnarray}
\braket{\vsigma'}{\vsigma}&=&\delta^3(\vsigma'-\vsigma), \\
\braketop{\vsigma'}{\hat{\mathtt{H}}_S}{\vsigma}&=&\hat{\mathtt{H}}_S(\vsigma')\delta^3(\vsigma'-\vsigma),\label{DiracOp}\\
\mathrm{and}\qquad\braketop{\vsigma}{\hat{\mathtt{H}}_S}{\psi}&=&\hat{\mathtt{H}}_S(\vsigma)\psi(\vsigma).
\end{eqnarray}
The formal solution for the time development of the fermionic wave
function can also be expressed with the star-exponential:
\begin{subequations}
\begin{eqnarray}
\psi(\vsigma,t)&=&e^{-\iu\hat{\mathtt{H}}_St}\psi(\vsigma) \\
        &=&\left(\cos(|E|t)-\iu\frac{\hat{\mathtt{H}}_S}{|E|}\sin(|E|t)\right)\psi(\vsigma) \\
        &=&\mathrm{Exp}_C(\mathtt{H}_St)\starC\psi(\vsigma).
\end{eqnarray}\label{TimeDev}
\end{subequations}

%%%%%%%%%%%%%%%%%%%%%%%%%%%%%%%%%%%%%%%%%%%%%%%%%%%%%%%%%%%%%%%%%%%%%%%%%%%%%%%%%%%%%%%%%%%%%%%%%%%%%
\section{Fermionic path integral}
\setcounter{equation}{0}
%%%%%%%%%%%%%%%%%%%%%%%%%%%%%%%%%%%%%%%%%%%%%%%%%%%%%%%%%%%%%%%%%%%%%%%%%%%%%%%%%%%%%%%%%%%%%%%%%%%%%
The fermionic path integral was formulated in \cite{Berezin}, where
the time development operator
$\hat{G}=e^{-\iu\hat{\mathtt{H}}_St}$ was expressed as a path
integral by using the integral form of the operator product. For
convenicence, we briefly demonstrate its derivation but in the context
of geometric algebra. The indices used here do not depict single
generators of a Grassmann algebra but whole sets of generators
($\vsigma_t=\{(\bsigma_t)_i\}$). As was shown in the last chapter,
the time development operator is simply the star exponential
\begin{equation}
\hat{G}\psi(\vsigma)=\mathrm{Exp}_C(\mathtt{H}_St)\starC\psi(\vsigma)
=G\starC\psi(\vsigma).
\end{equation}
This can also be expressed as an infinite product of star
exponentials
\begin{equation}
G=\lim_{N\rightarrow\infty}\left[\mathrm{Exp}_C(\mathtt{H}_St/N)\right]^{N\starC}
\end{equation}
and using the integral form of the star product (\ref{starCInt}) for
the time development (\ref{TimeDev}) one gets
\begin{equation}
G(\vsigma,t)=\lim_{N\rightarrow\infty}\int\prod_{m=1}^N
d^3\bsigma_md^3\bsigma'_m\exp\left(\sum_{n=1}^N\left(\vsigma_n\vsigma'_n+\vsigma'_n\vsigma_{n+1}+\vsigma_{n+1}\vsigma_n-\iu\mathtt{H}_S(\vsigma')t/N\right)\right)
\end{equation}
where $\vsigma_{N+1}=\vsigma$. To get a path integral formulation
similar to the bosonic case, the integration over
$\prod_{m=1}^Nd^3\bsigma_m$ has to be performed. Therefore, the
different sets of variables are separated by substituting
$\vsigma_n$ with $\vsigma''_n+\vsigma'''_n$ with new integration
variables $\vsigma''$, which results in the exponent
\begin{eqnarray}
&&\sum_{n=1}^{N-1}\left(\left(\vsigma''_n+\vsigma'''_n\right)\vsigma'_n+\vsigma'_n\left(\vsigma''_{n+1}+\vsigma'''_{n+1}\right)
+\left(\vsigma''_{n+1}+\vsigma'''_{n+1}\right)\left(\vsigma''_n+\vsigma'''_n\right)\right)\\
&&+\left(\vsigma''_N+\vsigma'''_N\right)\vsigma'_N+\vsigma'_N\vsigma+\vsigma\left(\vsigma''_N+\vsigma'''_N\right)\\
&=&\sum_{n=1}^{N-1}\left(\vsigma'_n\left(\vsigma'''_{n+1}-\vsigma'''_n\right)+\vsigma'''_{n+1}\vsigma'''_n\right)+\vsigma'_N\left(\vsigma-\vsigma'''_N\right)+\vsigma\vsigma'''_N\\
&&+\sum_{n=2}^{N-1}\left(\vsigma''_n\left(\vsigma'_n-\vsigma'_{n-1}-\vsigma'''_{n+1}+\vsigma'''_{n-1}\right)\right)
+\vsigma''_1\left(\vsigma'_1-\vsigma'''_2\right)+\vsigma''_N\left(\vsigma'''_{N-1}-\vsigma'_{N-1}+\vsigma'_N-\vsigma\right)\label{mixed}\\
&&+\sum_{n=1}^{N-1}\vsigma''_{n+1}\vsigma''_n,\label{Grauss}
\end{eqnarray}
where the summand with the Hamiltonian is omitted because it is not
altered. The mixed terms (\ref{mixed}) are eliminated by choosing
$\vsigma'''_{n+1}-\vsigma'''_{n-1}=\vsigma'_n-\vsigma'_{n-1}$ for
$n=2\ldots N-1$, $\vsigma'''_2=\vsigma'_1$, and
$\vsigma'''_{N-1}=\vsigma+\vsigma'_{N-1}-\vsigma'_N$. Thus, the rest
of the $\vsigma''$-terms (\ref{Grauss}) form the Grassmannian Gauss
integral and for even $N$ it yields
\begin{equation}
\int\prod_{m=1}^Nd^3\vsigma''_m\exp\left(\sum_{n=1}^{N-1}\vsigma''_{n+1}\vsigma''_n\right)=1.
\end{equation}
Collecting the remaining terms we get
\begin{eqnarray}
&&\sum_{n=1}^{N-1}\left(\vsigma'_n\left(\vsigma'''_{n+1}-\vsigma'''_n\right)+\vsigma'''_{n+1}\vsigma'''_n\right)+\vsigma'_N\left(\vsigma-\vsigma'''_N\right)+\vsigma\vsigma'''_N\\
&=&\sum_{n=1}^{N/2-1}\left(\vsigma'_{2n}\left(\vsigma'''_{2n+1}-\vsigma'''_{2n}\right)+\vsigma'''_{2n+1}\vsigma'''_{2n}
+\vsigma'_{2n+1}\left(\vsigma'''_{2n+2}-\vsigma'''_{2n+1}\right)+\vsigma'''_{2n+2}\vsigma'''_{2n+1}\right)\\
&&+\vsigma'_1\left(\vsigma'''_2-\vsigma'''_1\right)+\vsigma'''_2\vsigma'''_1+\vsigma'_N\left(\vsigma-\vsigma'''_N\right)+\vsigma\vsigma'''_N\\
&=&\sum_{n=1}^{N/2-1}\left(-\vsigma'_{2n}\vsigma'''_{2n}+\vsigma'_{2n+1}\vsigma'''_{2n+2}+\vsigma'''_{2n+1}\left(\vsigma'_{2n+1}-\vsigma'_{2n}\right)+\vsigma'''_{2n+1}\left(\vsigma'''_{2n}-\vsigma'''_{2n+2}\right)\right)\\
&&+\vsigma'_N\left(\vsigma-\vsigma'''_N\right)+\vsigma\vsigma'''_N\\
&=&\sum_{n=1}^{N/2}\vsigma'''_{2n}\left(\vsigma'_{2n}-\vsigma'_{2n-1}\right)+\vsigma\left(\vsigma'''_N-\vsigma'_N\right).
\end{eqnarray}
In the continual limit the substitution results in
$\vsigma'''(\tau)=\frac{1}{2}\left(\vsigma'(\tau)+\vsigma'(0)\right)$
so that the complete integral can then formally be written as
\begin{equation}
G(\vsigma,t)=\lim_{N\rightarrow\infty}\int\prod_{n=1}^N
d^3\bsigma'_n\exp\left(\int_0^t\left[\frac{1}{4}\vsigma'\dot\vsigma'-\iu\mathtt{H}_S(\vsigma')\right]d\tau+\frac{1}{2}\left(\vsigma\vsigma'_0+\vsigma'_0\vsigma'_t+\vsigma'_t\vsigma\right)\right),\label{pathintBerezin}
\end{equation}
where $\vsigma'_0=\vsigma'(0)$ and $\vsigma'_t=\vsigma'(t)$.

The Dirac notation, which was established in the previous
chapter, can be used for an alternative definition of the fermionic
Green's function.  Using (\ref{PsiDirac}) and (\ref{UnitDirac}) with
the formal solution for the time development yields
\begin{eqnarray}
\psi(\vsigma_t,t)&=&\braket{\vsigma_t}{\psi(t)} \\
    &=&\bra{\vsigma_t}e^{-\iu\hat{\mathtt{H}}_S(t-t_0)}\ket{\psi(t_0)} \\
    &=&\bra{\vsigma_t}e^{-\iu\hat{\mathtt{H}}_S(t-t_0)}\int d^3\bsigma_{t_0}\ket{\vsigma_{t_0}}\braket{\vsigma_{t_0}}{\psi(t_0)} \\
    &=&\int d^3\bsigma_{t_0}\bra{\vsigma_t}e^{-\iu\hat{\mathtt{H}}_S(t-t_0)}\ket{\vsigma_{t_0}}\psi(\vsigma_{t_0},t_0) \\
    &=&\int
    d^3\bsigma_{t_0}G(\vsigma_t,t;\vsigma_{t_0},t_0)\psi(\vsigma_{t_0},t_0).\label{Green}
\end{eqnarray}
So the fermionic Green's function is
\begin{equation}
G(\vsigma_t,t;\vsigma_{t_0},t_0)=\bra{\vsigma_t}e^{-\iu\hat{\mathtt{H}}_S(t-t_0)}\ket{\vsigma_{t_0}}.
\end{equation}
With (\ref{DiracOp}), (\ref{starCInt}), and (\ref{DeltaDef}) this
results in
\begin{eqnarray}
G(\vsigma_t,t;\vsigma_{t_0},t_0)&=&e^{-\iu\hat{\mathtt{H}}_S(\vssigma_t)(t-t_0)}\delta^3(\vsigma_t-\vsigma_{t_0})\\
&=&\mathrm{Exp}_C(\mathtt{H}_S(\vsigma_t)(t-t_0))\starC\delta^3(\vsigma_t-\vsigma_{t_0})\\
&=&\int
d^3\bsigma'd^3\bsigma''\mathrm{Exp}_C(\mathtt{H}_S(\vsigma')(t-t_0))\delta^3(\vsigma''-\vsigma_{t_0})e^{\vssigma_t\vssigma'+\vssigma'\vssigma''+\vssigma''\vssigma_t}\\
&=&-\int
d^3\bsigma'\mathrm{Exp}_C(\mathtt{H}_S(\vsigma')(t-t_0))e^{\vssigma_t\vssigma'+\vssigma'\vssigma_{t_0}+\vssigma_{t_0}\vssigma_t}.\label{starexpint1}
\end{eqnarray}
Substituting now $\vsigma'$ with $\vsigma_t-\vsigma'$ yields
\begin{eqnarray}
G(\vsigma_t,t;\vsigma_{t_0},t_0)&=&\int
d^3\bsigma'\mathrm{Exp}_C(\mathtt{H}_S(\vsigma_t-\vsigma')(t-t_0))e^{\vssigma'\vssigma_t+\vssigma_{t_0}\vssigma'}\\
&=&\int
d^3\bsigma'\mathrm{Exp}_C(\mathtt{H}_S(\vsigma_t-\vsigma')(t-t_0))e^{\vssigma'(\vssigma_t-\vssigma_{t_0})}.\label{starexpint2}
\end{eqnarray}
So, with the alternative definition of the Green's function one gets
a connection to the star exponential that is similar to the
connection in the bosonic case \cite{Cohen,Sharan}. The same result
can be obtained in another way. $G(\vsigma_t,t;\vsigma_{t_0},t_0)$
in (\ref{Green}) can be subdivided into infinitesimal components
\begin{eqnarray}
G(\vsigma_t,t;\vsigma_{t_0},t_0)&=&\lim_{N\rightarrow\infty}\int
\prod_{m=1}^{N}d^3\bsigma_{m}\prod_{n=N}^0G(\vsigma_{n+1},t_{n+1};\vsigma_n,t_n)\\
&=&\lim_{N\rightarrow\infty}\int
\prod_{m=1}^{N}d^3\bsigma_{m}\prod_{n=N}^0\mathrm{Exp}_C(\mathtt{H}_S(\vsigma_{n+1})(t_{n+1}-t_n))\starC\delta^3(\vsigma_{n+1}-\vsigma_n)
\end{eqnarray}
where $\vsigma_{N+1}=\vsigma_t$ and $\prod_{n=N}^0$ means that the
order of the factors is inverted to maintain the leading sign. 
(\ref{starexpint2}) can then be reobtained by using
again the integral form of the star product to evaluate the
expression and with an analogous substitution of the integrand and
comparison with (\ref{pathintBerezin}) and (\ref{starexpint1}).

%%%%%%%%%%%%%%%%%%%%%%%%%%%%%%%%%%%%%%%%%%%%%%%%%%%%%%%%%%%%%%%%%%%%%%%%%%%%%%%%%%%%%%%%%%%%%%%%%%%%%
\section{Conclusions}
\setcounter{equation}{0}
%%%%%%%%%%%%%%%%%%%%%%%%%%%%%%%%%%%%%%%%%%%%%%%%%%%%%%%%%%%%%%%%%%%%%%%%%%%%%%%%%%%%%%%%%%%%%%%%%%%%%
In this paper we have given an overview of several formalisms that
describe the spin with Grassmann variables. All these formulations
essentially lead back to the same algebra, namely the geometric
algebra ${\cal C}\ell_3^+(\mathbb{C})$ or alternatively ${\cal
C}\ell_3(\mathbb{R})$. To allow for a meaningful geometric
interpretation in the non-relativistic case at hand, one of these
two algebras with three Grassmann variables should be used as in
this case space is three-dimensional. The remaining issue is whether
the real algebra or the complex algebra is the better choice. The
real algebra is, at first, the obvious choice because it offers the
most elegant and plausible description of the spin whereas the
complex geometric algebra naturally appears in connection with the
bosonic star product formalism.  It remains to be seen which of the
formulations is better suited when extended to the relativistic case
using for example the spacetime algebra or the algebra of physical
space and when applied to specific problems.

%%%%%%%%%%%%%%%%%%%%%%%%%%%%%%%%%%%%%%%%%%%%%%%%%%%%%%%%%%%%%%%%%%%%%%%%%%%%%%%%%%%%%%%%%%%%%%%%%%%%%

\end{document}